\documentclass[conference]{IEEEtran}
\IEEEoverridecommandlockouts
\usepackage{cite}
\usepackage{amsmath,amssymb,amsfonts}
\usepackage{algorithmic}
\usepackage{graphicx}
\usepackage{subcaption}
\usepackage{colortbl}
\usepackage{multirow}
\usepackage{xcolor}
\usepackage{listings}
\usepackage{fancyvrb} 
\usepackage{soul}
\usepackage{array}
\usepackage{multirow}
\usepackage{todonotes}

\usepackage{float}
\usepackage{hyperref}
\def\BibTeX{{\rm B\kern-.05em{\sc i\kern-.025em b}\kern-.08em
    T\kern-.1667em\lower.7ex\hbox{E}\kern-.125emX}}
    
\begin{document}

\title{Generating Fake Cyber Threat Intelligence \\ Using Transformer-Based Models}

\author{\IEEEauthorblockN{Priyanka Ranade\IEEEauthorrefmark{1}, Aritran Piplai\IEEEauthorrefmark{1}, Sudip Mittal\IEEEauthorrefmark{2}, Anupam Joshi\IEEEauthorrefmark{1}, Tim Finin\IEEEauthorrefmark{1}, 
%, Mahmoud Abdelsalam\IEEEauthorrefmark{4}, Maanak Gupta\IEEEauthorrefmark{5}
}
%\IEEEauthorblockA{

\IEEEauthorrefmark{1}Department of Computer Science \& Electrical Engineering, University of Maryland, Baltimore County,\\Email: \{priyankaranade, apiplai1, joshi, finin\}@umbc.edu\\

\IEEEauthorrefmark{2}Department of Computer Science, University of North Carolina, Wilmington, \\ Email: mittals@uncw.edu\\

%\IEEEauthorrefmark{4} Dept. of Computer Science, Manhattan College, Email: mabdelsalam01@manhattan.edu\\
%\IEEEauthorrefmark{5} Dept. of Computer Science,
%Tennessee Technological University, Email: mgupta@tntech.edu\\
%\IEEEauthorrefmark{3} Cisco Research and Development, sand7@umbc.edu\\

}

\maketitle

\begin{abstract}
Cyber-defense systems are being developed to automatically ingest Cyber Threat Intelligence (CTI) that contains semi-structured data and/or text to populate knowledge graphs. A potential risk is that fake CTI can be generated and spread through Open-Source Intelligence (OSINT) communities or on the Web to effect a data poisoning attack on these systems. Adversaries can use fake CTI examples as training input to subvert cyber defense systems, forcing their models to learn incorrect inputs to serve the attackers' malicious needs. 

In this paper, we show how to automatically generate fake CTI text descriptions using transformers. Given an initial prompt sentence, a public language model like GPT-2 with fine-tuning can generate plausible CTI text that can mislead cyber-defense systems. We use the generated fake CTI text to perform a data poisoning attack on a Cybersecurity Knowledge Graph (CKG) and a cybersecurity corpus. The attack introduced adverse impacts such as returning incorrect reasoning outputs, representation poisoning, and corruption of other dependent AI-based cyber defense systems. We evaluate with traditional approaches and conduct a human evaluation study with cybersecurity professionals and threat hunters. % to test the accuracy and potential use-cases of the generated fake CTI. 
Based on the study, professional threat hunters were equally likely to consider our fake generated CTI and authentic CTI as true.
\end{abstract}


\section{Introduction}

Open-source platforms such as social media, the dark web, security blogs, and news sources play a vital role in providing the cybersecurity community with Cyber Threat Intelligence (CTI). This OSINT based threat intelligence complements sources collected by companies like IBM, Virtustotal or  Mandiant, by analyzing malware found in the wild, as well as that obtained by the Intelligence community. CTI is information about cybersecurity threats and threat actors that is shared with analysts and systems to help detect and mitigate harmful events. CTI can be shared as text or as semi-structured data with some text fields using formats like Structured Threat Information Expression (STIX) \cite{stix2} and Malware Information Sharing Platform (MISP) \cite{wagner2016misp}. Recent research has shown how text analysis approaches can be used to transform free text threat information into more structured forms\cite{cybertwitter,mittal2019cyber,mittal2017thinking,neil2018mining,ranade2018using,ranade2018understanding,samtani2020proactively,arnold2019dark,mulwad2011}, and even be ingested into policy driven defensive systems to enable detection\cite{narayanan2018early,patwardhan2004}. 

Although there are many clear benefits to open-source threat intelligence, addressing and handling \textit{misinformation} across these platforms is a growing concern. The misinformation risk for the security community is the possible dissemination of false CTI by threat actors in an attempt to poison systems that ingest and use the information\cite{khurana2019preventing}. In January 2021, Google Threat Analysis Group discovered an ongoing campaign that targets security researchers. Various nation state government-backed threat actors created fake accounts and blog posts with textual cybersecurity information on a variety of exploits in an attempt to divert security researchers from credible CTI sources \cite{fakecticampaign}. There is also additional research that suggests the possibility of future propagation of fake CTI. Maasberg et al. \cite{maasberg2018exploring} conducted a study of methods in propagating fake cybersecurity news and developed components to categorize it. The authors did not create fake cyber news, just studied its potential propagation. The widespread generation of fake CTI itself is heavily under-explored, and is a key contribution of this paper. 

The widespread propagation of fake CTI  primarily impacts cyber analysts who rely on the information to keep up to date with current attack vectors, as well as the cyber defense systems that ingest the information to take correct mitigation steps\cite{narayanan2018early}. Next-generation cyber defense systems are now being developed to automatically ingest and extract data from open source CTI to populate knowledge graphs, that are then used to detect potential attacks or as training data for machine learning systems.

Adversaries can use fake CTI as training input to subvert cyber defense systems. This type of attack is commonly known as a \textit{data poisoning attack} \cite{vorobeychik2018adversarial}. Many cyber defense systems that rely on this data automatically collect streams of CTI data from common sources. Adversaries can post fake CTI across open sources, infiltrating the training corpus of AI-based cyber defense systems with ease. This fake information will \textit{appear} legitimate to cyber analysts, but will in reality, have false components that contradict the real data. As can be seen from the examples in Table~\ref{tab:ctiexamples}, convincing fake CTI can be generated that provides incorrect information about the vulnerabilities exploited by an attack, or its consequences. This can cause confusion in analysts on what steps to take to address a threat. In an automated system cyber defense system that is ingesting the CTI, this can also break the reasoning and learning process altogether or force the model to learn incorrect inputs to serve the adversaries' malicious goals. Techniques demonstrated for open-source CTI can also be applied for covert data, such as proprietary information belonging to a particular company or government entity. In this scenario, potential attack strategies will more than likely be categorized as insider threats, and adversaries will be employees looking to exploit internal systems.

In this paper, we generate realistic fake CTI examples by fine-tuning the public GPT-2 model. Transformer-based methods are state-of-the art approaches that aid in detecting and generating misinformation on a large scale with minimal human effort \cite{grover2016node2vec}. 
Our generated fake CTI was able to confuse professional threat hunters and led them to label nearly all of the fake CTI as true. We also use the generated fake CTI examples to demonstrate data poisoning attacks on a Cybersecurity Knowledge Graph (CKG) and a cybersecurity corpus. We made sure that our generated fake data was never circulated in the wild, and remained on our machines where we generated it for testing.
%The generated results were evaluated through both a perplexity score and human study involving cybersecurity experts and threat hunters.

%\noindent
Our work makes three main contributions:
\begin{itemize}
  \item We produce a fine-tuned GPT-2 model that generates fake CTI text (Section \ref{section:finetuning}),
  \item We demonstrate a possible poisoning pipeline for infiltrating a CKG (Section \ref{section:poiosning}), and
  \item We present an evaluation and analysis of the fake and real CTI text (Sections \ref{section:generatingcti} and \ref{sec:eval}).
\end{itemize}

The organization of this paper is as follows - In Section \ref{section:backgroundandrelwork}, we present background and related work. We describe our fake CTI generation methodology in Section \ref{section:methodology}, which includes fine-tuning the GPT-2 transformer model on CTI data (Section \ref{section:finetuning}) and evaluating the generated fake CTI (Section \ref{sec:eval}). We showcase a data poisoning attack on a cybersecurity corpus and CKG (Section \ref{section:poiosning}) as well as provide additional experiments and analysis after ingesting the fake CTI with the CKG (Section \ref{section:ingestingFakeCTI}).
We conclude and present  future work in Section \ref{conc}.

\section{Background and Related Work}
\label{section:backgroundandrelwork}

In this section, we describe transformer architectures and related work in the areas of text generation, misinformation, AI-Based cybersecurity systems, knowledge graphs, and adversarial machine learning.

\subsection{Transformer Models
\label{section:transformermodels}}

Encoder-decoder configurations inspired current state-of-the art language models such as GPT \cite{radford2018improving} and BERT \cite{devlin2018bert} which utilize the transformer architecture \cite{vaswani2017attention}. Similar to Recurrent Neural Network (RNN) based sequence to sequence (Seq2Seq) models, the transformer encoder maps an input sequence into an abstract high dimensional space.  The decoder then transforms the vector into an output sequence. Unlike its Seq2Seq precursor, the transformer does not use any RNN components and relies solely on the attention mechanism to generate sequences. 

Seq2Seq architectures rely on LSTM cells to process an input sequence one word at a time. In a transformer model, all input words are processed in parallel. Due to this, the transformer introduces the concept of a \textit{positional encoding} in order to capture word ordering information in the n-dimensional vector of each word. The encoder and decoder components of the transformer also contain a multi-head attention mechanism. This can be described with the equation below where $Q$ represents queries, $K$ represents keys, and $V$ represents values. 

\vspace{-3mm}
\[\underbrace{\text{Attention}(Q,K,V)}_{\scriptstyle\text{Queries,Keys,Values}}=softmax\left(\dfrac{QK^T}{\sqrt{d_k}}\right){V}\mkern-45mu\]
\vspace{-3mm}

%Suppose $y$ is an input to the attention mechanism shown in the equation above \cite{Vaswani2017attention}. 
\noindent
The complete description of creating these values has been presented by Vaswani et al. \cite{vaswani2017attention}. At the start of the encoder, let $y$ be the initial sentence representation. As it travels through each layer of the encoder, $y$ gets updated by different encoder layers. %For example, encoder layer 2 receives an input $y$, which was the output $y$ of encoder layer 1. 
The input $y$ is used to calculate $Q$, $K$, and $V$ in the above equation. Attention is calculated by taking the transpose of the matrix dot product $QK$ and dividing by the square root of the dimension of the keys $\sqrt{d_k}$. Lastly, using the attention weights, we find the weighted sum of values $V$. The decoder attention mechanism operates similarly to the encoder, but employs \textit{masked multihead attention}. A linear and softmax layer are also added to produce the output probabilities of each word. In this paper, we focus on the GPT-2 model \cite{radford2019language} which exclusively uses decoder blocks.

\subsection{Transformer based Use-Cases % for Text Generation
\label{section:textgeneration}}

Generative transformer models have many use-cases such as machine translation \cite{wang2019learning}, question-answering \cite{shao2019transformer} and text summarization \cite{liu2019text}. A popular example of a generative transformer model is OpenAI GPT \cite{radford2018improving}. In recent years, GPT-2 \cite{radford2019language} and GPT-3 \cite{brown2020language,openaiapi} models have also been developed (At the time of writing this paper, GPT-3 is only accessible by a paywall API, and the model along with its other components are unavailable). GPT models across generations differ from each other in the sizes of data-sets used and number of parameters added. For example, the WebText dataset used to train GPT-2 contains eight million documents.% and 48 layers.%, while the BookCorpus dataset \cite{}, used to train GPT-1 contained 7000 documents and 12 layers. 

In this paper, we utilize GPT-2 in our experiments. Unlabeled data is used to \textit{pretrain} an unsupervised GPT model for a generic task. \textit{Fine-tuning} the generic pre-trained models is a common method of extending the architectures for more specific tasks \cite{radford2018improving}. Lee et al. \cite{lee2019patent} produced patent claims by fine-tuning the generic pretrained GPT-2 model with U.S. utility patents claims data. Similarly, Feng et al. \cite{feng2020genaug} fine-tuned GPT-2 on a small set of yelp review data-set and used it as a baseline model for various augmentation experiments. %They chose the yelp review data because it differed substantially from the WebText corpus. 
%\hl{Another novel method of improving text generation for downstream tasks involves combining transformers with knowledge graphs. Koncel-Kedziorski et al. \cite{koncel2019text} create a graph transforming encoder to generate multi-sentence texts from the output of a knowledge graph. }

%\subsection{AI and Misinformation
%\label{section:misinformation}} 
Transformers have been utilized to both detect and generate \textit{misinformation}. Misinformation can be generally categorized as lies, fabricated information, unsupported facts, misunderstandings, and outdated facts and is often used to achieve economic, political, or social gain \cite{del2016spreading}. %There is substantial recent evidence of utilizing transformers in the area of misinformation.
Vijjali et al. \cite{vijjali2020two} utilize BERT-based transformers to detect false claims surrounding the COVID-19 pandemic. Similarly, Zellers et al. \cite{zellers2019defending} also use a BERT-based model called Grover, which can detect and generate neural fake news. Their evaluation shows that human beings found machine-generated disinformation more trustworthy than human-written information. 

%\subsection{Cyber Threat Intelligence
%\label{section:cti}}

\begin{figure*}[t] % move as necessary for good final placement
  \centering
  \includegraphics[scale=0.63]{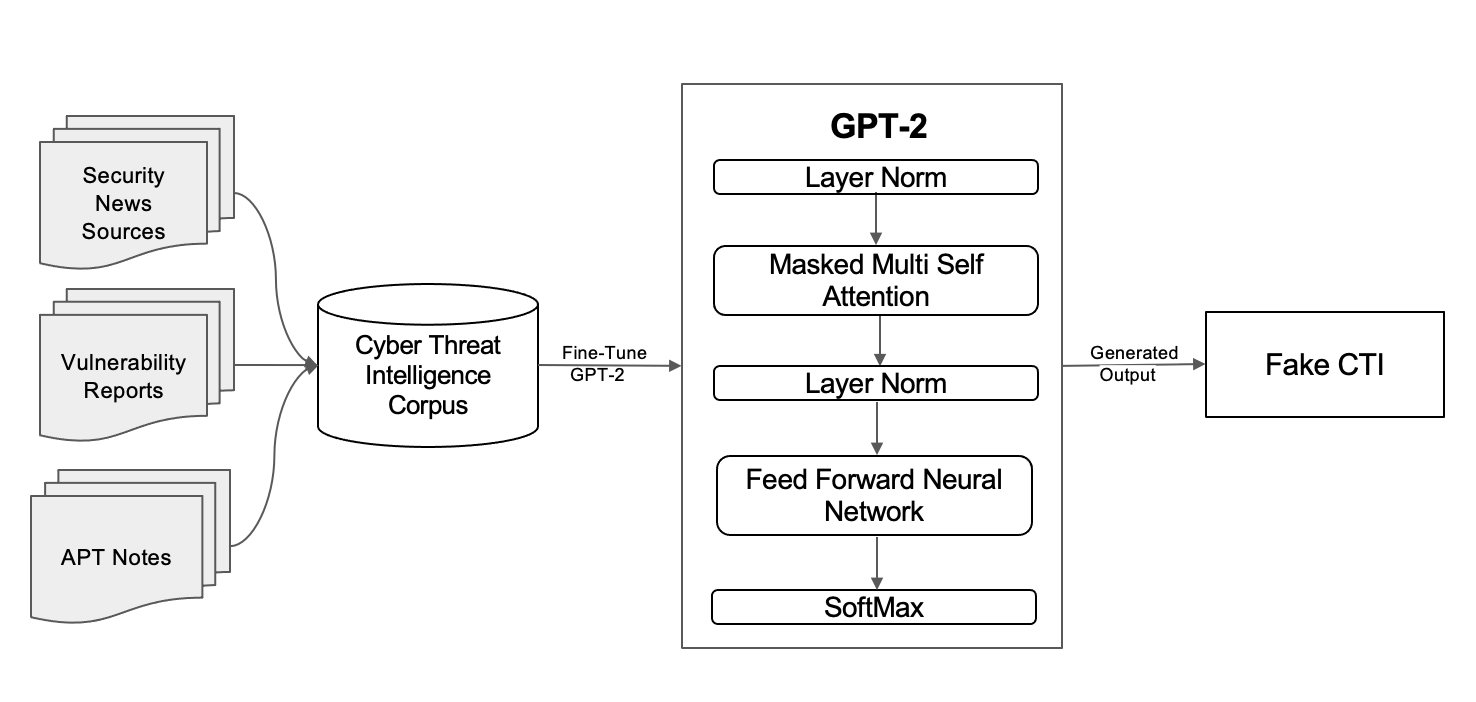}
  \caption{We collected cybersecurity-related text from several OSINT sources and used it to fine-tune the public GPT-2 model, which generated fake CTI descriptions.}
  \label{image:system_architecture}
  \vspace{-4mm}
\end{figure*}

\subsection{AI-Based Cyber Systems and Knowledge Graphs
\label{section:aibasedcyber}}

Next-generation cyber defense systems use various knowledge representation techniques such as word embeddings and knowledge graphs in order to improve system inference on potential attacks. The use of CTI is an integral component of such systems. Knowledge graphs for cybersecurity have been used before to represent various entities \cite{pinglerelext,piplai2020creating,gao2021system}. Open source CTI has been used to build Cybersecurity Knowledge Graphs (CKG) and other agents to aid cybersecurity analysts working in an organization \cite{cybertwitter,mittal2019cyber,mittal2017thinking,neil2018mining,ranade2018using,ranade2018understanding,samtani2020proactively,arnold2019dark}. Mittal et al. created Cyber-All-Intel and CyberTwitter \cite{mittal2017thinking,cybertwitter} which utilizes a variety of knowledge representations such as a CKG to augment and store CTI. %CyberTwitter \cite{cybertwitter} is another example of a framework that automatically issues vulnerability alerts to users by converting vulnerability information contained in tweets to RDF representation and using the Unified Cybersecurity Ontology (UCO) to provide the system with cybersecurity domain information.

The use of knowledge graphs for cyber-defense tasks has also been used in malware analysis tasks \cite{liumalware,parkmalware,blakemalware,joshi2016alda,joshi2017semantically}. Piplai et al. \cite{piplai2020creating,piplaibehavior} create a pipeline to extract information from malware after action reports and other unstructured CTI sources and represent that in a CKG. %They also improve this CKG by using malware behavior data to add more granular information about malware to the CKG \cite{piplaibehavior}. 
They use this prior knowledge stored in a CKG as input to agents in a reinforcement learning environment %. Using the stored intelligence, the agents were able to learn malware detection strategies after detonating malware samples 
\cite{piplaiusing2020}. We demonstrate the effects of the poisoning attack, by ingesting fake CTI on CKG using a complete CTI processing pipeline \cite{piplai2020creating,pinglerelext}. %In our previous work, we also investigated multilingual CTI available and developed an encoder-decoder based translation engine to produce domain-specific translations for cyber analysts \cite{ranade2018using,ranade2018understanding}. There are also examples of AI-based offensive systems. 
%ybersecurity Knowledge Graphs have been widely used to represent Cyber Threat Intelligence (CTI). We use CKG to store CTI as semantic triples that helps in understanding how different cyber entities are related. This representation allows the users to query the system and reason over the information. %The classes and possible relationships between their instances are dictated by a pre-defined schema. This schema also helps in reasoning, as it can help define specific constraints and axioms that may exist between the classes. %For example, we can define a class `A' disjoint from a class `B'. So, if we assert an entity in class `A' and `B' both, the knowledge graph reasoner is going to display an error and not let us perform the action.
 %,,khurana2019preventing, piplai2019creating}. 

%CKGs have also been used to compare different malware by Liu et al. \cite{liumalware}. Some behavioral aspects have also been incorporated in CKGs, where the authors used system call information \cite{parkmalware}. Graph based methods have also been post processed by machine learning algorithms, as demonstrated by other approaches \cite{blakemalware,joshi2016alda,joshi2017semantically}.  

\subsection{Adversarial Machine Learning and Poisoning Attacks
\label{section:poisoning attacks}}

Adversarial machine learning is a technique used to subvert machine learning systems by providing deceptive inputs to their models. Adversaries use these methods to manipulate AI-based system learning in order to alter protected behavior and serve their own malicious goals \cite{joseph2019adversarial}. There are several types of adversarial techniques such as evasion, functional extraction, inversion, and poisoning attacks \cite{vorobeychik2018adversarial}. In this paper, we focus on \textit{data poisoning attack} strategies. 

Data poisoning attacks directly compromise the integrity of an AI system that uses machine learning by contaminating its training data \cite{barreno2006can,rubinstein2009antidote,kloft2010online,kloft2012security}. These methods rely heavily on the use of synthesized and/or incorrect input data. AI-based cyber defense systems can potentially include fake data into their training corpus. The attacker dominates future output by ensuring the system learns fake inputs and performs poorly on actual data. Biggio et al. \cite{biggio2012poisoning} demonstrated pioneering methods in using kernelized gradient ascent strategies to produce malicious input that can be used to predict future decisions of a support vector machine.

In recent years, poisoning attacks have grown to target cyber-defense systems. One such attack is the VirusTotal poisoning attack demonstrated by the McAfee Advanced Threat Research team \cite{datapoisoningexamples}. This attack compromised several intrusion detection systems that ingest VirusTotal data. The attacker created mutant variants of a ransomware family sample and uploaded the mutants to the VirusTotal platform. Intrusion detection systems that ingest VirusTotal data classified the mutant files as the particular ransomware family. %The scope of this attack also targeted AI-based cyber defense systems that use the data-set to classify malware families. 
Similarly, Khurana et al. perform credibility checks on incoming CTI. They develop a reputation score that is used by systems and analysts to evaluate the level of trust for input intelligence data \cite{khurana2019preventing}. Duddu et al. survey several methods of using machine learning to model adversary behavior \cite{duddu2018survey}.

%Types of AML
%USE THIS   - %https://github.com/mitre/advmlthreatmatrix/blob/master/pages/adversarial-ml-101.md#adversarial-machine-learning-101

%example 
%https://github.com/mitre/advmlthreatmatrix/blob/master/pages/case-studies-page.md#case-studies-page
%VirusTotal Poisoning

%other examples from the link
%Maasberg et al. \cite{Maasberg2018} conduct a study of methods in propagating fake cybersecurity news. They formulate primary differences between fake and actual cyber news as well as develop components to categorize different types of fake cyber news. The authors in their paper do not create fake cyber news and only study its propagation. 

\section{Methodology}
\label{section:methodology}

 In this section we describe our fake CTI generation pipeline. Figure \ref{image:system_architecture}, presents the overall approach. We begin by creating a cybersecurity corpus in Section \ref{sec:corpuscreation}. The cybersecurity corpus contains a collection of CTI from a variety of OSINT sources. We then fine-tune the pre-trained GPT-2 model on our cybersecurity corpus (Section \ref{section:finetuning}). The fine-tuned model allows us to automatically generate large collections of fake CTI samples. We then evaluate our model and describe a poisoning attack against a CTI extraction pipeline. % (See Section \ref{sec:eval} \& \ref{section:ingestingcti}). % We then utilize the generated fake cyber threat intelligence examples to poison an AI-based cyber defense system . 

\subsection{Creating a Cybersecurity Corpus
\label{sec:corpuscreation}}

We categorize our CTI collection into three main sources, as shown in Figure \ref{image:system_architecture}. We collect security news articles, vulnerability databases, and technical Advanced Persistent Threat (APT) reports. The security news category contains 1000 articles from Krebs on Security \cite{krebs}. The vulnerability reports contain 16,000 Common Vulnerability and Exposures (CVE) records provided by MITRE Corporation and National Vulnerability Database (NVD) from years 2019-2020 \cite{nvd2013}. Lastly, we collect 500 technical reports on APTs from the available APTNotes repository \cite{aptnotesrepo}. %, and 500 arXiv cybersecurity research paper documents \cite{rahman18}. 

The widespread use of the above sources across the greater security community establishes our corpus as a gold standard for cybersecurity domain information. Security news articles are common sources used by cybersecurity threat hunters to stay current on the latest vulnerabilities and exploits. In particular, \textit{Krebs on Security} is a global resource utilized and referenced by the Security Operations Centers (SOCs) and popular security bloggers. The resource is updated nearly daily with reports describing exploits having medium to high impact that security analysts and companies have found. \textit{APT Reports} is a repository of documents written by malware analysts and includes fine-grained technical briefings of advanced persistent threat groups and persistent malware strains. The \textit{CVE database}, maintained by MITRE Corporation, is another example of fine-grained OSINT and is used as a common resource for corporations to track vulnerabilities and exploits associated with popular products they produce and use. By including both general and fine-grained OSINT, we can fine-tune the GPT-2 to learn about various facets of the security community that are otherwise not present in the GPT-2's training data, derived from a collection of web pages. More on the GPT-2 fine-tuning process, is available in the next section.

%In the next section, we discuss methods used to fine-tune the GPT-2 model with the cybersecurity corpus described above.

\begin{table*}[t]
\centering
%\refstepcounter{table}
\begin{tabular}{|l|l|c|} 
\hline
\multicolumn{1}{|c|}{\textbf{Generation Input} }                                                                                 & \multicolumn{1}{c|}{\textbf{Cyber Threat Intelligence} }                                                                                                                                                                                                                                                                                                                                                                                                                                                 & \textbf{True/Fake}                                                                       \\ 
\hline
\multirow{2}{*}{\begin{tabular}[c]{@{}l@{}} APT41 is a \\ state-sponsored \\ espionage group \end{tabular}}                      & {\cellcolor[rgb]{0.906,0.906,0.906}}\begin{tabular}[c]{@{}>{\cellcolor[rgb]{0.906,0.906,0.906}}l@{}}APT41 is a state-sponsored espionage group which operates against higher education, travel\\~services, and news/media firms which provide some indication that the group also tracks \\individuals and conducts surveillance.\end{tabular}                                                                                                                                                           & {\cellcolor[rgb]{0.906,0.906,0.906}}\textbf{True CTI}                                    \\ 
\cline{2-3}
                                                                                                                                 & \begin{tabular}[c]{@{}l@{}}APT41 is a state-sponsored espionage group. The Russian, Moscow-based group is responsible for \\capturing and stealing computer network communications and intellectual property. APT41 \\employs social engineering techniques to get around network defenses designed to last for \\years, and is capable of compromising and stealing dozens of companies\end{tabular}                                                                                                             & \begin{tabular}[c]{@{}c@{}}\textbf{Fake CTI }\\\textbf{ (generated)} \end{tabular}  \\ 
\hline
\multirow{2}{*}{\begin{tabular}[c]{@{}l@{}} Malicious Domain \\ in SolarWinds \\ Hack Turned \\ into ‘Killswitch' \end{tabular}} & {\cellcolor[rgb]{0.906,0.906,0.906}}\begin{tabular}[c]{@{}>{\cellcolor[rgb]{0.906,0.906,0.906}}l@{}}Malicious Domain in SolarWinds Hack Turned into ‘Killswitch'. A key malicious domain name \\used to control potentially thousands of computer systems compromised via the months-long \\breach at network monitoring software vendor SolarWinds was commandeered by security \\experts and used as a “killswitch” designed to turn the sprawling cybercrime operation \\against itself.\end{tabular} & {\cellcolor[rgb]{0.906,0.906,0.906}}\textbf{True CTI}                                    \\ 
\cline{2-3}
                                                                                                                                 & \begin{tabular}[c]{@{}l@{}}Malicious Domain in SolarWinds Hack Turned into Killswitch Service. This attack took place \\last year and affected hundreds of organizations. The attackers took advantage of the\\~vulnerability in Win32k framework to attack their targets through crafted code.\end{tabular}                                                                                                                                                                                             & \begin{tabular}[c]{@{}c@{}}\textbf{Fake CTI }\\\textbf{ (generated)} \end{tabular}  \\ 
\hline
\multirow{2}{*}{\begin{tabular}[c]{@{}l@{}} An issue was \\ discovered in the \\ Quiz and Survey \\ Master plugin \end{tabular}} & {\cellcolor[rgb]{0.906,0.906,0.906}}\begin{tabular}[c]{@{}>{\cellcolor[rgb]{0.906,0.906,0.906}}l@{}}An issue was discovered in the Quiz and Survey Master plugin before 7.0.1 for WordPress. \\ It made it possible for unauthenticated attackers to upload arbitrary files and achieve \\ remote code execution.\end{tabular}                                                                                                                     & {\cellcolor[rgb]{0.906,0.906,0.906}}\textbf{True CTI}                                    \\ 
\cline{2-3}
                                                                                                                                 & \begin{tabular}[c]{@{}l@{}}An issue was discovered in the Quiz and Survey Master plugin 1.2.1 for WordPress. By \\setting a CSRF token, an attacker can take advantage of a vulnerability in the survey.php \\admin panel.\end{tabular}                                                                                                                                                                                                                                                                  & \begin{tabular}[c]{@{}c@{}}\textbf{Fake CTI }\\\textbf{ (generated)} \end{tabular}  \\
\hline
\end{tabular}
\caption{Fake CTI Samples produced by our fine-tuned GPT-2 model.}
\label{tab:ctiexamples}
\vspace{-4mm}
\end{table*}

\subsection{Fine-Tuning GPT-2 on Cyber Threat Intelligence Data \label{section:finetuning}}
 
The original GPT-2 model was trained with the WebText dataset \cite{radford2019language} collected from eight million web pages. While this dataset contains some general cybersecurity text, it lacks much of the fine-grained CTI information useful to the security community. To address this problem, we fine-tuned the general model with the cybersecurity corpus described above. The diverse CTI sources in our corpus gives the GPT-2 model a variety of examples and the ability to adapt to several aspects of the cybersecurity domain. %\hl{Our cybersecurity corpus contains fine-grained CTI information which is unique to the security community and is not present in general news sources.} 
\textit{Pre-trained} transformer-based language models like GPT-2 are easily adapted to new domains such as cybersecurity. Instead of training from scratch and initializing with random weights, we start with the model with pre-trained parameters. We used the publicly released pre-trained GPT-2 model with 117M parameters which has 12 layers, 786 dimensional states, and 12 attention heads.

During training, we divide the corpus in a 35\% train and test split. We set block size as 128, batch size as 64, and learning rate as 0.0001. We utilize the Gaussian Error Linear Unit (GELU) activation function. The GPT-2 architecture shown in Figure\ref{image:system_architecture}, consists of normalization layers \cite{ba2016layer}, an attention layer, a standard feed forward neural network, and a soft-max layer. The feed forward neural network contains 786*4 dimensions. We trained the model for twenty three hours (20 epochs) and achieved a perplexity value 35.9. Examples of the generated CTI and more details on our experimentation are given in the next section. %\ref{section:generatingcti}}.

%As discussed in Section \ref{section:transformermodels}, the GPT-2 solely consist of decoder blocks. The inputs are passed through a normalization layer, then through the first block of the attention layer. The block outputs are once again passed to a normalization layer and fed to a feed forward neural network that accepts outputs from the attention layer and adds an activation function and dropout. Its output is passed through a softmax layer, which obtains the positional encoding of the highest probability word inside the vocabulary. 

\subsection{Generating Fake CTI}
\label{section:generatingcti}

We use our fine-tuned GPT-2 model to generate fake CTI examples, three of which are shown in Table \ref{tab:ctiexamples}. The generation process is initiated with a prompt that is provided as input to the fine-tuned GPT-2 model (the first column in Table \ref{tab:ctiexamples}). The model uses the prompt to generate the fake CTI. The generation process is shown in Figure \ref{image:system_architecture}. The tokenized prompt is passed through a normalization layer, then through the first block of the attention layer. The block outputs are also passed to a normalization layer and fed to a feed forward neural network, which adds an activation function and dropout. Its output is passed through a softmax layer, which obtains the positional encoding of the highest probability word inside the vocabulary. 

%We provide an example for each of the three data categories present in our cybersecurity corpus described in Section \ref{sec:corpuscreation}. \hl{The table shows both the true CTI and the fake version of the CTI generated by our fine-tuned model (Column 2), given a CTI starter sentence as input (Column 1).} The fake examples consist of topics that are related to each other, but not relevant to the main input topic we feed into the model. 

The \textit{first sample} in Table \ref{tab:ctiexamples}, provides information on APT group APT41. Given the prompt, \textit{``APT41 is a state sponsored espionage group"}, the model was able to form a partially false narrative about APT41. APT41 is a Chinese state-sponsored espionage group, not a Russian group as indicated by the model. Although this is a false fact, the later part of the generated CTI is partially true. Despite some true information, the incorrect nation-state information surrounding APT41 is still present and adds conflicting intelligence if ingested by an AI-based cyber defense system. 

In the \textit{second example}, we provide an input prompt from a Krebs on Security article \cite{krebssolarwinds}. The model generated fake CTI, which states \textit{kill switch} as an actual service, when in actuality, \textit{kill switch} refers to the method of disconnecting networks from the Internet. In addition, it relates the false service to the \textit{Win32k} framework. This gives the fake CTI enough credibility and seems true to cyber analysts. %It can also potentially mislead an AI-based cyber defense system as shown in Section \ref{section:ingestingcti}. %We show an example of poisoning rules in an AI-based cyber defense system in Section \ref{section:ingestingcti}.

Lastly for the \textit{third example}, we provide an input prompt from a 2019 CVE record. The model generated the correct product, but an incorrect associated version and attack type; the true attack was a remote code execution while the generated attack was privilege escalation. While a remote code execution attack can be related to a privilege escalation attack in general, the specific context of using a Cross-Site Request Forgery (CSRF) token to gain access to  survey.php is incorrect for this specific product.

\subsection{Evaluating the generated CTI}\label{sec:eval}

We next show that the generated fake CTIs are credible. We use two approaches to show this. First, we evaluate the ability of the fine-tuned model to predict our test data by calculating the perplexity score. Next, we conduct human evaluation studies. The study required a group of cybersecurity professionals and threat hunters to label a collection of generated and actual CTI samples as true or fake. The cybersecurity experience of the participants range from 2-30 years (in operational settings), with an average experience of 15 years. The idea is to see if professionals in the field can separate real CTI from fake instances generated by our system. 

%We evaluate the generated fake CTI examples using perplexity scores and a human . Perplexity measures the language models' ability to predict test data. To determine how well the model performs under a real-world task, we we conduct a human evaluation study. 

%In this section we describe our methods for evaluating the quality and efficacy of generated text. We define “quality” as a metric to measure the similarity of machine-generated text to human-generated text. We measure quality through perplexity scores and human-evaluation. Components such as sentence structure and grammar are addressed. We define “efficacy” as a metric to measure legitimacy of the generated text. We determine sentence-level similarities between generated CTI and the actual CTI. We measure efficacy with the BertScore metric. Lastly, we conduct a study among a group of cybersecurity professionals to classify generated fake CTI as true or fake. 

 %As the model adapted to cybersecurity data, the loss value steadily decreased (See Figure \ref{fig:loss}). %Our initial experiments overfitted the model to the cybersecurity corpus. To avoid overfitting, we adjusted the learning rate after every 100 steps using a learning rate scheduler. 

 In the context of cybersecurity, human evaluation with potential real-world users of the fake CTI is more indicative than traditional methods such as perplexity scores. The main objective of generating fake CTI is to mislead cyber analysts and bypass intelligence pipelines that they frequently monitor. If the generated CTI does not possess a high range of malformed sentence structure, poor grammar, or incomprehensible text (obvious mistakes indicating the text was produced by a machine), we can assume it has fair potential to appear real to analysts. Perplexity is a common method to determine ``uncertainty'' in a language model, by assigning probabilities to the test set. Perplexity is measured as the exponentiated average logarithmic loss and ranges from 0-100. The lower the perplexity score, the less uncertainty exists within the model. The base 117M GPT-2 model we fine-tuned has a perplexity score of 24 \cite{lee2019patent}. %Other state-of-the-art models they generated using smaller datasets achieved perplexity scores ranging from 20-30. 
We ensure the model is not evaluated on text from the training set by calculating perplexity on a separate test set and achieve a calculated perplexity score of 35.9, showing strong ability of the model to generate plausible text.  

%\todo{what does this mean though ? Is this a good score? Compared to what? We need some more discussion to put this in context}

In order to evaluate the potential implications of the generated fake CTI in a real world setting, we conduct a study across a group of ten cybersecurity professionals and threat hunters\footnote{Our study protocol was evaluated by UMBC's IRB and classfied as Not Human Subjects Research}. We provided the participants with an assessment set of both true and fake CTI text samples. Using their own expertise, participants labeled each text sample in the corpus as either true or fake. We created the assessment set by collecting 112 text samples of true CTI drawn from various sources described in Section \ref{sec:corpuscreation}. We pre-process the text samples by truncating them to the first 500 words and eliminating partial last sentences. We select the first sentence of each sample as an initial prompt to the fine-tuned GPT-2 model and generate a fake CTI example of no more than 500 words. We further divide the 112 samples (56 true CTI and their generated fake counterparts) into two separate annotation sets to ensure true CTI and direct fake counterparts are not part of the same annotation task. Therefore, each annotation task included 28 samples of true text and 28 non-overlapping samples of generated fake data. We randomize the data in each annotation task assigned to the participants.% and provide each half of the participants with a different test set.

Participants worked individually, and labeled each of the 56 samples as either true or fake. Participants used their own judgement in labeling each sample, and were prohibited to use external sources like search engines during the assessment. The results of the study are provided in the confusion matrix. 

The confusion matrix shows the true positive, false negative, false positive, and true negative rates for 560 CTI samples (including both true and fake data). Of the total 560 samples that were rated, the accuracy (36.8\%) was less than chance. The threat hunters predicted 52.5\% incorrectly (74 true samples as false and 220 false statements as true) and 47.5\% samples correctly (206 true samples as true and 60 false statements as false). Despite their expertise, the threat hunters were only able to label 60/280 of the generated samples as fake and found the a large majority (78.5\%) of the fake samples as true. These results demonstrate the ability of the generated CTI to confuse security experts, and portends trouble if such techniques are widely used.% in real world scenarios. 

\begin{center}
\newcommand\MyBox[2]{
  \fbox{\lower0.75cm
    \vbox to 1.2cm{\vfil
      \hbox to 1.2cm{\hfil\parbox{1.4cm}{#1\\#2}\hfil}
      \vfil}%
  }%
}
\noindent
\renewcommand\arraystretch{1.5}
\setlength\tabcolsep{0pt}
\begin{tabular}{c >{\bfseries}r @{\hspace{0.7em}}c @{\hspace{0.4em}}c @{\hspace{0.7em}}l}
  \multirow{10}{*}{\rotatebox{90}{\parbox{1.1cm}{\bfseries\centering Actual Data \newline}}} & 
    & \multicolumn{2}{c}{\bfseries Participant Labels} & \\
  & & \bfseries True & \bfseries False & \bfseries Total \\
  & True & \MyBox{206}{Samples}& \MyBox{74} {Samples} & 280 \\[2.4em]
  & False & \MyBox{220}{Samples} & \MyBox{60}{Samples} & 280 \\
  & Total & 426 & 134 &
\end{tabular}
\end{center}

We further investigated the fake samples that were accurately labeled as fake and observed more linguistic errors in the text than in comparison to the fake samples that were labeled as true. Although the majority of the fake CTI contained entities (such as products and attack vectors) that were unrelated to each other, we found if the sentence structure displayed little or no linguistic deficiencies, the data was likely labeled as true. We also noticed sources that lacked substantial context were likely labeled as false. 

The generated fake CTI not only has the ability to mislead cybersecurity professionals, but also has the ability to infiltrate cyber defense systems. In the next section, we describe how the generated fake CTI examples can be used to launch a data poisoning attack.

\section{Data Poisoning using Fake CTI}
\label{section:poiosning}

With the fake CTI examples in Table \ref{tab:ctiexamples} we can easily simulate a \textit{data poisoning attack} where the fake CTI is used as training input to subvert knowledge extraction pipelines such as those described by Piplai et al. \cite{piplai2020creating}, Mittal et al. \cite{mittal2019cyber,cybertwitter}, Gao et al. \cite{gao2021system,gao2021system2}, and Arnold et al. \cite{arnold2019dark}. Here an attacker can skillfully position fake CTI on multiple OSINT sources like Twitter, Stack Overflow, dark web forums, and blogs.

%Aritran please fix these figures
\begin{figure}[ht!]
    \centering
    \includegraphics[width=\columnwidth]{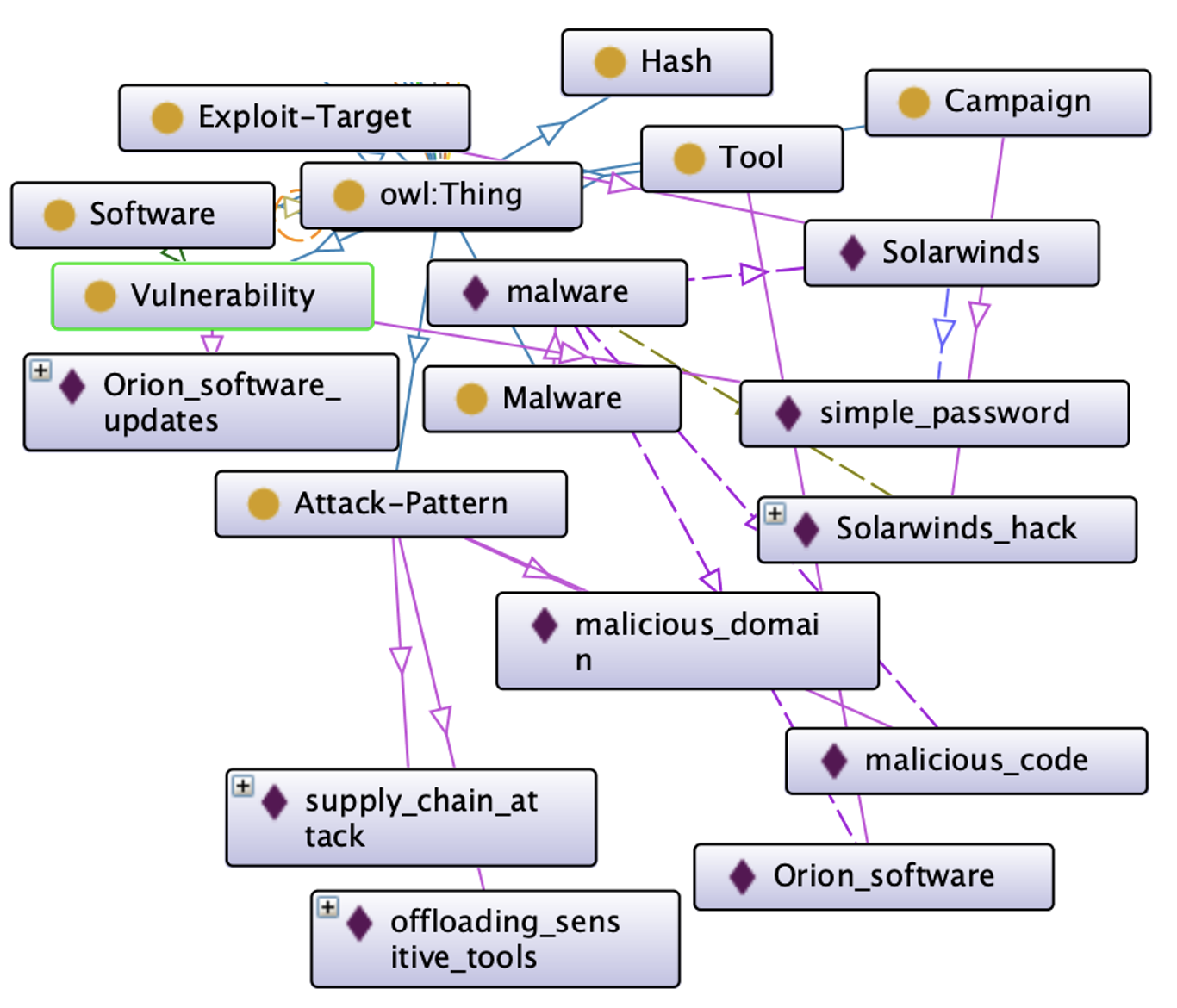}
    \caption{CKG populated with data from legitimate true CTI sources.}
    \label{image:CKGoutputreal}
    \vspace{-4mm}
\end{figure}

\begin{figure}[ht!]
    \centering
    \includegraphics[width=\columnwidth]{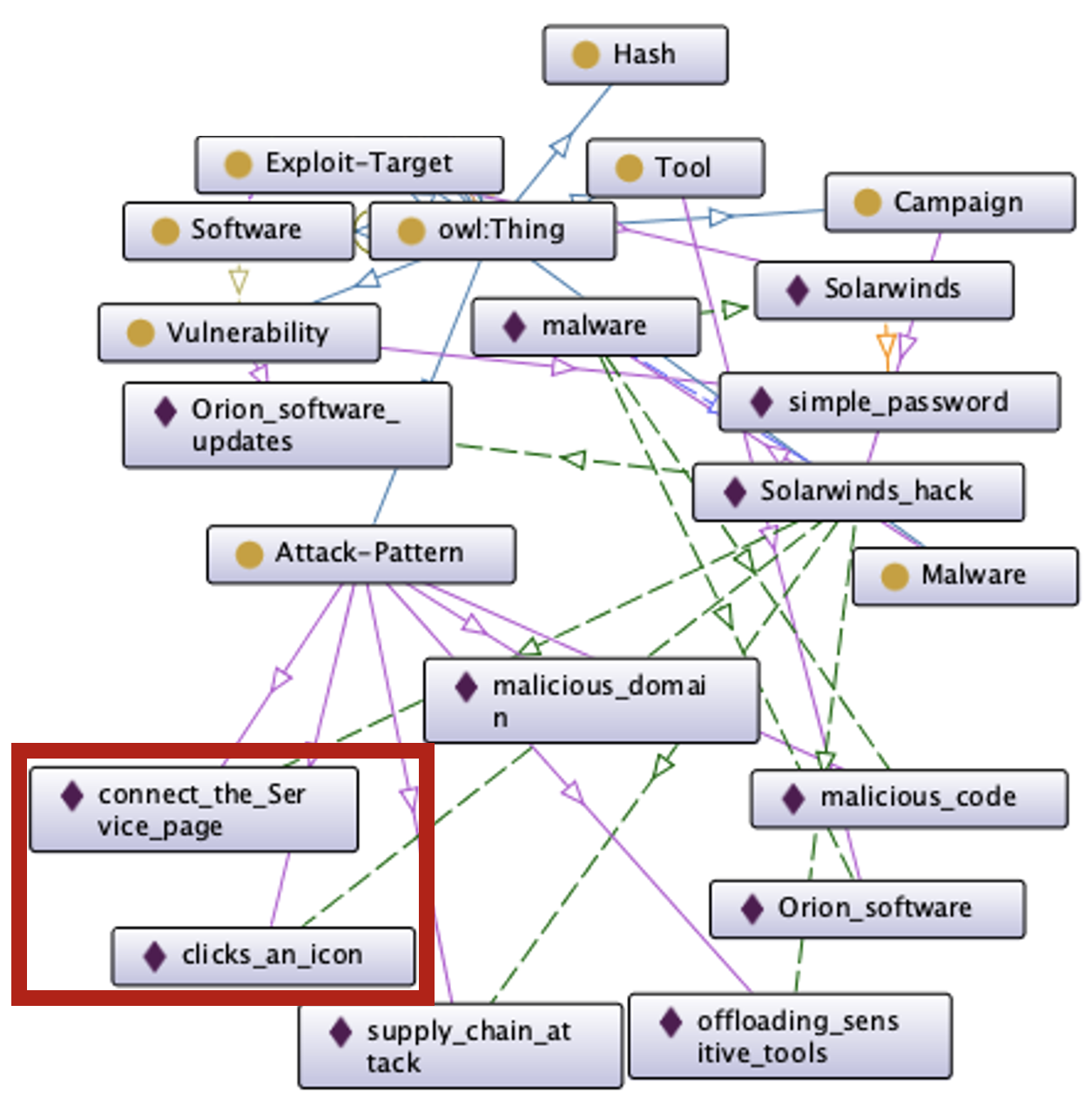}
    \caption{The poisoned CKG with additional data (red box) extracted from fake CTI.}
    \label{image:CKGoutputfake}
    \vspace{-4mm}
\end{figure}

Many of the systems described above include native crawlers along with cybersecurity concept extractors, entity relationship extractors, and knowledge representation techniques such as word embeddings, tensors, and knowledge graphs. These either use keyword-based methodologies or depend on AI tools to collect and process the CTI. Many of these systems can be easily tricked into including the fake CTI data in a \textit{cybersecurity corpus} along with the true CTI. This is especially possible if the attacker is able to craft the fake CTI in such a way that it ``appears very similar'' to true CTI. This fake information will then be ingested by a knowledge extraction pipeline utilized to create knowledge representations like, Cybersecurity Knowledge Graphs (CKG). %Ingesting fake CTI can adversely impact the reasoning and learning process altogether in order serve the adversaries' goals. 
%These corpora have generally been used to train AI models that help process input CTI. . 
Poisoning a corpus with fake CTI can enable an attacker to contaminate the training data of various AI systems in order to obtain a desired outcome at inference time. With influence over the CTI training data, an attacker can guide the creation of AI models, where an arbitrary input will result in a particular output useful to the attacker.

%In Section \ref{pipeline}, 
Next, we describe an attack on a popular knowledge representation technique that involves a CKG \cite{piplai2020creating, mittal2019cyber,pinglerelext}. As we already have access to a complete CTI processing pipeline that outputs a CKG \cite{piplai2020creating}, we choose to demonstrate the effects of the poisoning attack on the CKG. Once the fake CTI has been represented in a knowledge representation it can be used to influence other AI systems that depend on these representations. We also discuss the effects of the poisoning attack on the CKG in Section \ref{ckgeffect}.   %The model could be "reprogrammed" to perform a new undesired task. Further, access to training data would allow the attacker to create an offline model and create a Model Evasion. 	

%\section{Experiments and Analysis}

%\section{Ingesting fake CTI}
%\label{section:ingestingcti}

%The objective of creating a structured knowledge graph from the unstructured CTI text is to aid security professionals in their research. The security professionals can look up past knowledge about cyber-incidents, perform reasoning, and retrieve information with the help of queries. However, if generated fake information is ingested by the CKG, it can have detrimental impact on the quality of knowledge present in the CKG. We will demonstrate these effects in the following sections, where a fake CTI sample is processed upon by a cybersecurity concept extractor, relationship extractor and asserted into the CKG. 
%produce erroneous results to queries made by security professionals

%architecture diagram

%take fake intel put in CKG and embeddings.

\subsection{Processing fake CTI} \label{pipeline}
A CTI ingestion pipeline described in Piplai et al. \cite{piplai2020creating} and similar systems \cite{arnold2019dark,gao2021system,gao2021system2} take a CTI source as an input and produces a CKG as an output. The CKG contains cyber entities and their existing relationships. The first stage is a cybersecurity \textit{concept extractor} that takes a CTI and extracts various cyber entities. This is done by using a Named Entity Recognizer (NER) trained on a cybersecurity corpus. The second stage, is a deep-neural network based \textit{relationship extractor} that takes word embeddings of cyber entity pairs as an input and identifies likely relationships. This results in an entity-relationship set that can be asserted into the CKG. As a running example, we use the following \textit{fake} CTI text as input to the extraction pipeline-

\begin{quote}
\textit{`Malicious domain in SolarWinds hack turned into killswitch service where the malicious user clicks an icon (i.e., a cross-domain link) to connect the service page to a specific target.'}
\end{quote}

%\subsubsection{Output of a Cybersecurity Concept Extractor} 
%The cybersecurity concept extractor is a NER trained on a cybersecurity corpus. In the past our research group, has worked on developing cybersecurity concept extractors \cite{cybertwitter,piplai2020creating}. The existing concept extractor is based on Conditional Random Fields (CRFs) and Gibbs' sampling. The task is structured as a classification problem and the model fits each word to a set of entity classes defined by our CKG schema. The CKG schema is based on UCO \cite{syed2016uco,pinglerelext} and STIX \cite{stix2}. %We also use regular expressions to detect entities that do not need contextual information for prediction and can be identified by their pattern.
%An example of how the Concept Extractor works on a fake CTI is as follows.
When fake CTI is ingested by the pipeline, the cybersecurity concept extractor will output classifications that serve the adversaries' goals.
%We poison the CKG with the generated fake CTI. Our data poisoning attack is described below.
The concept extractor classifies \textit{`clicks an icon', `connect the service'} as `Attack-Pattern'. It also classifies \textit{`SolarWinds hack'} as a `Campaign'. These entities are extracted from the fake CTI potentially poisoning the CKG. 
%\subsubsection{Output of a Relationship Extractor}
%The Relationship Extractor \cite{pinglerelext} is a neural network that takes a pair of entity embeddings and produces a relationship as an output, if it exists. The list of legitimate relationships that can exist between pairs of entities is dictated by the CKG schema. 

The relationship extractor while processing the fake CTI above, outputs the following relationships:
\begin{itemize}
    \item `Solarwinds hack' (Campaign)-\textit{uses}-
    `clicks an icon' (Attack-Pattern).
    \item `Solarwinds hack' (Campaign)- \textit{uses} - `connect the service' (Attack-Pattern).
\end{itemize}

%\subsubsection{Asserting knowledge in the CKG}
The extracted entity relationship set can then be asserted in the CKG. %If we encounter similar entities from multiple sources we fuse them with the help of the `owl:SameAs' property. More information about this pipeline can be found in the paper by Piplai et al. \cite{piplai2020creating}. 
Figures \ref{image:CKGoutputreal} and \ref{image:CKGoutputfake}, describe the state of the CKG \textit{before} and \textit{after} asserting knowledge extracted from fake CTI. Figure \ref{image:CKGoutputreal}, contains entities and relationships extracted from true CTI samples describing the campaign `SolarWinds hack'. We can see entities like `Orion Software', identified as `Tool', and `malicious code' identified as `Attack-Pattern'. These entities are used by the malware in the `SolarWinds hack' and are present in the true CTI. We also see `simple password' as a vulnerability. Figure \ref{image:CKGoutputfake}, contains additional information extracted from fake CTI generated by our model. These additional entities and relationships have been asserted along with the entity `SolarWinds hack', and are demarcated by the red box. In this figure, we can see additional `Attack-Patterns' like, `connect the service page' and `clicks an icon' being captured in the CKG. These entities have been extracted using the pipeline from the \textit{fake} CTI and are an evidence of how a poisoned corpus with fake CTI can be ingested and represented in a CKG.
%\cite{piplai2020creating}

%\subsubsection{Generating graph embeddings} add  to section 3D IF INCLUDING

\subsection{Effects of fake CTI ingestion} \label{ckgeffect}
\label{section:ingestingFakeCTI}
The objective of creating a structured knowledge graph from the unstructured CTI text is to aid security professionals in their research. The security professionals can look up past knowledge about cyber incidents, perform reasoning, and retrieve information with the help of queries. 
%Other use-cases include designing network/system rules to identify and mitigate future attacks. 
However, if generated fake information is ingested by the CKG as part of a data poisoning attack, it can have detrimental impacts such as returning wrong reasoning outputs, bad security alert generation, representation poisoning, model corruption, etc.

%The primary objective of a CKG is to provide security professionals with contextual knowledge about an attack. Security professionals can retrieve data with the help of queries and also design rules to identify and mitigate future attacks. 
%\hl{A data poisoning attack  The presence of entities and relationships extracted from fake CTIs can have adverse effects on these use-cases. Next, we discuss some effects of this data poisoning scenario like, wrong reasoning outputs, and it's }

%\subsubsection{Wrong reasoning outputs} 
%If a security professional wants to execute queries to retrieve information about a cyber-attack, knowledge from fake CTI sources, can result in incorrect results. 
For example, if a security professional is interested in knowing which attack campaigns have used `click-baits', they will be misled by the result `Solarwinds hack'. As the fake CTI has been ingested and represented in the knowledge representation (See Section \ref{pipeline}). The following SPARQL \cite{sparql} query when executed on the CKG,
\begin{Verbatim}[fontsize=\small]
  SELECT ?x WHERE {
    ?x a CKG:Campaign;
       CKG:uses CKG:clicks_an_icon.}
\end{Verbatim}
%\end{lstlisting}
\noindent
will result in the following value:
\begin{Verbatim}[fontsize=\small]
  Solarwinds_hack
\end{Verbatim}
If security professionals are interested to know more information about `Solarwinds-hack', they may also receive incorrect information after executing appropriate SPARQL queries. % querychanged to use  inverse property operator ^
\begin{Verbatim}[fontsize=\small]
  SELECT ?x WHERE {
    ?x a CKG:Attack-Pattern;
       ^CKG:uses CKG:Solarwinds-hack.}
\end{Verbatim}
%\end{lstlisting}
\noindent
This query results in the following values:
\begin{Verbatim}[fontsize=\small]
  malicious_code, offloading_sensitive_tools, 
  connect_the_service_page, clicks_an_icon
\end{Verbatim}

Although we obtained some true results (sourced from true CTI), the presence of fake CTI guided results like, `connect the service page' and `clicks an icon' have the potential to mislead security professionals. %and sideline them from developing effective mitigation strategies. % by poisoning their research resources. %Although some of the results obtained are true, the majority are not and are a result of fake CTI ingestion. %\subsubsection{Wrong alert generation} 
Security professionals model cybersecurity attacks and generate network/system detection rules using past available information on the same attacks or similar attacks. They also use these representations to generate alerts for future attacks. For example, a `supply chain attack' exploiting a `small password' vulnerability `offloading sensitive tools' may mean that a new variant of the SolarWinds hack has surfaced. However, if prior knowledge contains fake CTI about the same attack, incorrect alerts can be generated.

More concerning, is the possibility of adversaries further optimizing the generated fake CTI to achieve more sophisticated and targeted changes to a CKG. One approach is to include a second stage to the fake CTI generation, by replacing entities such as IP addresses or process names, with targeted entities chosen by the adversary. This will cause the changes to be populated into the CKG, and the adversary can manipulate the system to treat the chosen entities as benign. After extracting a knowledge graph of the generated text, entities can be identified and replaced to look consistent with actual CTI sources. In this case the attacker can leverage various knowledge provenance methods, which augment the fake CTI knowledge graph with actual source information. These strategies can further confuse cyber defense professionals. We are exploring these more targeted attacks in ongoing future work.

%Using an information extraction system to capture the text and string offsets of the entity mentions the adversary can automatically populate targeted concepts into a CKG. 

%`Clicking an icon' and `connecting to the service page' have the potential to be part of a misleading rule that generates an alert for a future variant of the SolarWinds hack. %Such alerts can be incorrect as the rules have been constructed with incorrect information from fake CTIs. 
%\subsubsection{Miscellaneous effects} \more
%\hl{training other AI systems}

%A poisoned CTI corpus will also effect other cybersecurity representations like knowledge graphs, word embeddings, tensors, etc.

Once these knowledge representations are poisoned, additional defense systems can also be adversely impacted by fake cybersecurity information. For example, many of the insights generated by knowledge graphs are useful to other systems like AI-based intrusion detection systems \cite{kumar2010use,blakemalware,parkmalware}, or alert-generators \cite{cybertwitter,gao2021system}, reaching a larger breadth of linked systems and cybersecurity professionals.

%Coinciding use of fake information can lead to mass propagation of fake cybersecurity information across many sources, reaching a larger breadth of linked systems and cybersecurity professionals. 

%Fake CTI could then be shared on the Web or embedded in STIX and MISP objects, which could result in misinformation being added to a CTI knowledge graph or database.

\section{Conclusion \& Future Work}\label{conc}

%using transformers and later used the generated text to poison AI-based cyber defense systems such as a Cybersecurity Knowledge Graph (CKG).

In this paper, we automatically generated fake CTI text descriptions by fine-tuning the GPT-2 transformer using a cybersecurity corpus rich in CTI sources. %These CTI sources include cybersecurity news, vulnerability databases, and APT reports. 
By fine-tuning the GPT-2 transformer with cybersecurity text, we were able to adapt the general model to the cybersecurity domain. Given an initial prompt, the fine-tuned model is able to generate realistic fake CTI text examples. %We evaluate our model with perplexity scores and human evaluation. 
Our evaluation with cybersecurity professionals shows that generated fake CTI could easily mislead cybersecurity experts. We found that cybersecurity professionals and threat hunters labeled the majority of the fake CTI samples as true despite their expertise, showing that they found the fake CTI samples believable.

We use the fake CTI generated by the fine-tuned GPT-2 model to demonstrate a data poisoning attack on a knowledge extraction system that automatically ingests open sourced CTI. %Fake CTI can be shared on the Web or embedded in STIX and MISP objects, which could result in misinformation being added to a CTI knowledge graph or database.  
We exemplify the impacts of ingesting fake CTI, by comparing the state of the CKG before and after the data poisoning attack. 
%After the poisoning attack the CKG included entities and relations present in the fake CTI. %The CKG representing the true CTI extracted relevant entities and asserted correct relationships, while the poisoned CKG extracted additional entities leading to incorrect entities and relationships asserted in the CKG. 
The adverse impacts of these fake CTI sourced assertions include wrong reasoning outputs, representation poisoning, and model corruption.

In ongoing work, we are exploring defences against such data poisoning attacks. One approach is to develop systems that can detect linguistic errors and disfluencies that generative transformers commonly produce, but humans rarely make. %For example, we found participants in our study recognized fake text through grammatical errors and/or broken sentence structures. We may be able to train a classifier to identify linguistic inaccuracies in text and determine if it was likely to have been produced by a transformer rather than written by a person. 
%\hl{Our future work will explore how to defend against such data poisoning attacks. One approach is to develop systems that can detect the kinds of linguistic errors and disfluencies that generative transformers can produce but people rarely make. For example, we noticed that our GPT-2 model will sometimes generate text with the same sentence twice in a row. Another example of unnatural text we observed was an overly long list of countries in which an attack had been observed that contained several duplicates.}
A second approach to detecting fake CTI text can use a combination of novelty, consistency, provenance, and trust. CTI sources can be given a score that indicates the amount of trust the user wishes to include in their information. 

%CTI information from well known and reliable sources like CERT, NIST, or MITRE might be assumed to be true. CTI from less trusted open web sources/dark web that contains novel facts requires some evaluation. CTI with facts that contradicts or is inconsistent with trusted information we already know is likely, but not certain, to be fake.

\section*{Acknowledgement}

This work was supported by a U.S. Department of Defense grant, a gift from IBM research, and National Science Foundation grant \#2025685. We would like to thank various cybersecurity professionals and threat hunters at US defense contractors that took part in our human evaluation study. 

\bibliographystyle{unsrt}
\bibliography{bibliography/references}

\begin{thebibliography}{10}

\bibitem{stix2}
Oasis group.
\newblock Stix 2.0 documentation.
\newblock \url{https://oasis-open.github.io/cti-documentation/stix/}, May 2013.

\bibitem{wagner2016misp}
Cynthia Wagner, Alexandre Dulaunoy, G{\'e}rard Wagener, and Andras Iklody.
\newblock Misp: The design and implementation of a collaborative threat
  intelligence sharing platform.
\newblock In {\em Workshop on Information Sharing and Collaborative Security},
  pages 49--56. ACM, 2016.

\bibitem{cybertwitter}
Sudip Mittal, Prajit Das, Varish Mulwad, Anupam Joshi, and Tim Finin.
\newblock Cybertwitter: Using twitter to generate alerts for cybersecurity
  threats and vulnerabilities.
\newblock {\em IEEE/ACM Int. Conf. on Advances in Social Networks Analysis and
  Mining}, pages 860--867, 2016.

\bibitem{mittal2019cyber}
Sudip Mittal, Anupam Joshi, and Tim Finin.
\newblock Cyber-all-intel: An {AI} for security related threat intelligence.
\newblock {\em arXiv:1905.02895}, 2019.

\bibitem{mittal2017thinking}
Sudip Mittal, Anupam Joshi, and Tim Finin.
\newblock Thinking, fast and slow: Combining vector spaces and knowledge
  graphs.
\newblock {\em arXiv:1708.03310}, 2017.

\bibitem{neil2018mining}
Lorenzo Neil, Sudip Mittal, and Anupam Joshi.
\newblock Mining threat intelligence about open-source projects and libraries
  from code repository issues and bug reports.
\newblock In {\em Intelligence and Security Informatics}. IEEE, 2018.

\bibitem{ranade2018using}
Priyanka Ranade, Sudip Mittal, Anupam Joshi, and Karuna Joshi.
\newblock Using deep neural networks to translate multi-lingual threat
  intelligence.
\newblock In {\em International Conference on Intelligence and Security
  Informatics}, pages 238--243. IEEE, 2018.

\bibitem{ranade2018understanding}
Priyanka Ranade, Sudip Mittal, Anupam Joshi, and Karuna~Pande Joshi.
\newblock Understanding multi-lingual threat intelligence for {AI} based
  cyber-defense systems.
\newblock In {\em IEEE International Symposium on Technologies for Homeland
  Security}, 2018.

\bibitem{samtani2020proactively}
Sagar Samtani, Hongyi Zhu, and Hsinchun Chen.
\newblock Proactively identifying emerging hacker threats from the dark web: A
  diachronic graph embedding framework (d-gef).
\newblock {\em Transactions on Privacy and Security}, 23(4):1--33, 2020.

\bibitem{arnold2019dark}
Nolan Arnold, Mohammadreza Ebrahimi, Ning Zhang, Ben Lazarine, Mark Patton,
  Hsinchun Chen, and Sagar Samtani.
\newblock Dark-net ecosystem cyber-threat intelligence (cti) tool.
\newblock In {\em International Conference on Intelligence and Security
  Informatics}, pages 92--97. IEEE, 2019.

\bibitem{mulwad2011}
Varish Mulwad, Wenjia Li, Anupam Joshi, Tim Finin, and Krishnamurthy
  Viswanathan.
\newblock Extracting information about security vulnerabilities from web text.
\newblock In {\em 2011 IEEE/WIC/ACM International Conferences on Web
  Intelligence and Intelligent Agent Technology}, volume~3, pages 257--260,
  2011.

\bibitem{narayanan2018early}
Sandeep Narayanan, Ashwini Ganesan, Karuna Joshi, Tim Oates, Anupam Joshi, and
  Tim Finin.
\newblock Early detection of cybersecurity threats using collaborative
  cognition.
\newblock In {\em 4th Int. Conf. on Collaboration and Internet Computing},
  pages 354--363. {IEEE}, 2018.

\bibitem{patwardhan2004}
A.~Patwardhan, V.~Korolev, L.~Kagal, and A.~Joshi.
\newblock Enforcing policies in pervasive environments.
\newblock In {\em The First Annual International Conference on Mobile and
  Ubiquitous Systems: Networking and Services, 2004. MOBIQUITOUS 2004.}, pages
  299--308, 2004.

\bibitem{khurana2019preventing}
Nitika Khurana, Sudip Mittal, Aritran Piplai, and Anupam Joshi.
\newblock Preventing poisoning attacks on {AI} based threat intelligence
  systems.
\newblock In {\em 29th Int. Workshop on Machine Learning for Signal
  Processing}, pages 1--6. IEEE, 2019.

\bibitem{fakecticampaign}
Google Threat~Analysis Group.
\newblock {New campaign targeting security researchers}.
\newblock
  https://blog.google/threat-analysis-group/new--campaign-targeting-security-researchers/,
  2021.

\bibitem{maasberg2018exploring}
Michele Maasberg, Emmanuel Ayaburi, Charles Liu, and Yoris Au.
\newblock Exploring the propagation of fake cyber news: An experimental
  approach.
\newblock In {\em 51st Hawaii International Conference on System Sciences},
  2018.

\bibitem{vorobeychik2018adversarial}
Yevgeniy Vorobeychik and Murat Kantarcioglu.
\newblock Adversarial machine learning.
\newblock {\em Synthesis Lectures on Artificial Intelligence and Machine
  Learning}, 12(3):1--169, 2018.

\bibitem{grover2016node2vec}
Aditya Grover and Jure Leskovec.
\newblock node2vec: Scalable feature learning for networks.
\newblock In {\em 22nd ACM SIGKDD international conference on Knowledge
  discovery and data mining}, pages 855--864, 2016.

\bibitem{radford2018improving}
Alec Radford, Karthik Narasimhan, Tim Salimans, and Ilya Sutskever.
\newblock Improving language understanding by generative pre-training.
\newblock Technical report, {OpenAI}, 2018.

\bibitem{devlin2018bert}
Jacob Devlin, Ming-Wei Chang, Kenton Lee, and Kristina Toutanova.
\newblock Bert: Pre-training of deep bidirectional transformers for language
  understanding.
\newblock {\em arXiv:1810.04805}, 2018.

\bibitem{vaswani2017attention}
Ashish Vaswani, Noam Shazeer, Niki Parmar, Jakob Uszkoreit, Llion Jones,
  Aidan~N Gomez, {\L}ukasz Kaiser, and Illia Polosukhin.
\newblock Attention is all you need.
\newblock In {\em Advances in neural information processing systems}, pages
  5998--6008, 2017.

\bibitem{radford2019language}
Alec Radford, Jeffrey Wu, Rewon Child, David Luan, Dario Amodei, and Ilya
  Sutskever.
\newblock Language models are unsupervised multitask learners.
\newblock {\em OpenAI blog}, 1(8):9, 2019.

\bibitem{wang2019learning}
Qiang Wang, Bei Li, Tong Xiao, Jingbo Zhu, Changliang Li, Derek~F Wong, and
  Lidia~S Chao.
\newblock Learning deep transformer models for machine translation.
\newblock {\em arXiv:1906.01787}, 2019.

\bibitem{shao2019transformer}
Taihua Shao, Yupu Guo, Honghui Chen, and Zepeng Hao.
\newblock Transformer-based neural network for answer selection in question
  answering.
\newblock {\em IEEE Access}, 7:26146--26156, 2019.

\bibitem{liu2019text}
Yang Liu and Mirella Lapata.
\newblock Text summarization with pretrained encoders.
\newblock In {\em Conf. on Empirical Methods in Natural Language Processing and
  the 9th Int. Joint Conf. on Natural Language Processing}, pages 3721--3731.
  ACL, 2019.

\bibitem{brown2020language}
Tom~B Brown, Benjamin Mann, Nick Ryder, Melanie Subbiah, Jared Kaplan, Prafulla
  Dhariwal, Arvind Neelakantan, Pranav Shyam, Girish Sastry, and Amanda Askell.
\newblock Language models are few-shot learners.
\newblock {\em arXiv:2005.14165}, 2020.

\bibitem{openaiapi}
{OpenAI}.
\newblock {Open AI API}.
\newblock \url{https://openai.com/blog/openai-api/}, 2021.

\bibitem{lee2019patent}
Jieh-Sheng Lee and Jieh Hsiang.
\newblock Patent claim generation by fine-tuning {OpenAI} {GPT-2}.
\newblock {\em arXiv:1907.02052}, 2019.

\bibitem{feng2020genaug}
Steven~Y Feng, Varun Gangal, Dongyeop Kang, Teruko Mitamura, and Eduard Hovy.
\newblock Genaug: Data augmentation for finetuning text generators.
\newblock In {\em Deep Learning Inside Out: 1st Workshop on Knowledge
  Extraction and Integration for Deep Learning Architectures}, pages 29--42,
  2020.

\bibitem{del2016spreading}
Michela Del~Vicario, Alessandro Bessi, Fabiana Zollo, Fabio Petroni, Antonio
  Scala, Guido Caldarelli, H~Eugene Stanley, and Walter Quattrociocchi.
\newblock The spreading of misinformation online.
\newblock {\em Proceedings of the National Academy of Sciences},
  113(3):554--559, 2016.

\bibitem{vijjali2020two}
Rutvik Vijjali, Prathyush Potluri, Siddharth Kumar, and Sundeep Teki.
\newblock Two stage transformer model for {COVID-19} fake news detection and
  fact checking.
\newblock {\em arXiv:2011.13253}, 2020.

\bibitem{zellers2019defending}
Rowan Zellers, Ari Holtzman, Hannah Rashkin, Yonatan Bisk, Ali Farhadi,
  Franziska Roesner, and Yejin Choi.
\newblock Defending against neural fake news.
\newblock In {\em Advances in neural information processing systems}, pages
  9054--9065, 2019.

\bibitem{pinglerelext}
Aditya Pingle, Aritran Piplai, Sudip Mittal, Anupam Joshi, James Holt, and
  Richard Zak.
\newblock Relext: Relation extraction using deep learning approaches for
  cybersecurity knowledge graph improvement.
\newblock {\em IEEE/ACM International Conference on Advances in Social Networks
  Analysis and Mining}, 2019.

\bibitem{piplai2020creating}
Aritran Piplai, Sudip Mittal, Anupam Joshi, Tim Finin, James Holt, and Richard
  Zak.
\newblock Creating cybersecurity knowledge graphs from malware after action
  reports.
\newblock {\em IEEE Access}, 8:211691--211703, 2020.

\bibitem{gao2021system}
Peng Gao, Xiaoyuan Liu, Edward Choi, Bhavna Soman, Chinmaya Mishra, Kate
  Farris, and Dawn Song.
\newblock A system for automated open-source threat intelligence gathering and
  management.
\newblock {\em arXiv preprint arXiv:2101.07769}, 2021.

\bibitem{liumalware}
Jing Liu, Yuan Wang, and Yongjun Wang.
\newblock The similarity analysis of malicious software.
\newblock In {\em Int. Conf. on Data Science in Cyberspace}. IEEE, 2016.

\bibitem{parkmalware}
Younghee Park, Douglas Reeves, Vikram Mulukutla, and Balaji Sundaravel.
\newblock Fast malware classification by automated behavioral graph matching.
\newblock In {\em 6th Annual Workshop on Cyber Security and Information
  Intelligence Research}. ACM, 2010.

\bibitem{blakemalware}
Blake Anderson, Daniel Quist, Joshua Neil, Curtis Storlie, and Terran Lane.
\newblock Graph-based malware detection using dynamic analysis.
\newblock {\em Journal in Computer Virology}, 7(1):247--258, 2011.

\bibitem{joshi2016alda}
Karuna~P Joshi, Aditi Gupta, Sudip Mittal, Claudia Pearce, Anupam Joshi, and
  Tim Finin.
\newblock Alda: Cognitive assistant for legal document analytics.
\newblock In {\em AAAI Fall Symposium}, 2016.

\bibitem{joshi2017semantically}
Maithilee Joshi, Sudip Mittal, Karuna~P Joshi, and Tim Finin.
\newblock Semantically rich, oblivious access control using {ABAC} for secure
  cloud storage.
\newblock In {\em Int. Conf. on edge computing}, pages 142--149. IEEE, 2017.

\bibitem{piplaibehavior}
Aritran Piplai, Sudip Mittal, Mahmoud Abdelsalam, Maanak Gupta, Anupam Joshi,
  and Tim Finin.
\newblock Knowledge enrichment by fusing representations for malware threat
  intelligence and behavior.
\newblock In {\em International Conference on Intelligence and Security
  Informatics}. IEEE, 2020.

\bibitem{piplaiusing2020}
Aritran Piplai, Priyanka Ranade, Anantaa Kotal, Sudip Mittal, Sandeep
  Narayanan, and Anupam Joshi.
\newblock {Using Knowledge Graphs and Reinforcement Learning for Malware
  Analysis}.
\newblock In {\em 4th International Workshop on Big Data Analytics for Cyber
  Intelligence and Defense, IEEE International Conference on Big Data}. IEEE,
  December 2020.

\bibitem{joseph2019adversarial}
Anthony~D Joseph, Blaine Nelson, Benjamin~IP Rubinstein, and JD~Tygar.
\newblock {\em Adversarial Machine Learning}.
\newblock Cambridge University Press, 2019.

\bibitem{barreno2006can}
Marco Barreno, Blaine Nelson, Russell Sears, Anthony~D Joseph, and J.~Doug
  Tygar.
\newblock Can machine learning be secure?
\newblock In {\em ACM Symposium on Information, computer and communications
  security}, pages 16--25, 2006.

\bibitem{rubinstein2009antidote}
Benjamin Rubinstein, Blaine Nelson, Ling Huang, Anthony Joseph, Shing-hon Lau,
  Satish Rao, Nina Taft, and J.~Doug Tygar.
\newblock Antidote: understanding and defending against poisoning of anomaly
  detectors.
\newblock In {\em ACM SIGCOMM Conference on Internet Measurement}, pages 1--14,
  2009.

\bibitem{kloft2010online}
Marius Kloft and Pavel Laskov.
\newblock Online anomaly detection under adversarial impact.
\newblock In {\em Proceedings of the Thirteenth International Conference on
  Artificial Intelligence and Statistics}, pages 405--412. JMLR Workshop and
  Conference Proceedings, 2010.

\bibitem{kloft2012security}
Marius Kloft and Pavel Laskov.
\newblock Security analysis of online centroid anomaly detection.
\newblock {\em The Journal of Machine Learning Research}, 13(1):3681--3724,
  2012.

\bibitem{biggio2012poisoning}
Battista Biggio, Blaine Nelson, and Pavel Laskov.
\newblock Poisoning attacks against support vector machines.
\newblock {\em arXiv preprint arXiv:1206.6389}, 2012.

\bibitem{datapoisoningexamples}
MITRE.
\newblock {Virus Total Data Poisoning Case Studies}.
\newblock
  http://git\-hub\-.com/mitre/advmlthreatmatrix/blob/master/pages/case-studies-page.\-md\#virustotal-poisoning,
  2021.

\bibitem{duddu2018survey}
Vasisht Duddu.
\newblock A survey of adversarial machine learning in cyber warfare.
\newblock {\em Defence Science Journal}, 68(4), 2018.

\bibitem{krebs}
Brian Krebs.
\newblock Krebs on security.
\newblock \url{https://krebsonsecurity.com/}, 2021.

\bibitem{nvd2013}
Harold Booth, Doug Rike, and Gregory Witte.
\newblock The national vulnerability database (nvd): Overview.
\newblock Technical report, National Institute of Standards and Technology,
  2013.

\bibitem{aptnotesrepo}
aptnotes.
\newblock {APTnotes repository}.
\newblock \url{https://github.com/aptnotes/data}, 2021.

\bibitem{ba2016layer}
Jimmy~Lei Ba, Jamie~Ryan Kiros, and Geoffrey~E Hinton.
\newblock Layer normalization.
\newblock {\em stat}, 1050:21, 2016.

\bibitem{krebssolarwinds}
Brian Krebs.
\newblock {Malicious Domain in Solarwinds Hack turned into killswitch}.
\newblock
  https://krebsonsecurity.com/2020/12/malicious-domain-in-solarwinds-hack-turned-into-killswitch/,
  2021.

\bibitem{gao2021system2}
Peng Gao, Fei Shao, Xiaoyuan Liu, Xusheng Xiao, Haoyuan Liu, Zheng Qin,
  Fengyuan Xu, Prateek Mittal, Sanjeev~R Kulkarni, and Dawn Song.
\newblock A system for efficiently hunting for cyber threats in computer
  systems using threat intelligence.
\newblock {\em arXiv preprint arXiv:2101.06761}, 2021.

\bibitem{sparql}
W3.
\newblock Sparql query language.
\newblock \url{https://www.w3.org/TR/rdf-sparql-query/}.

\bibitem{kumar2010use}
Gulshan Kumar, Krishan Kumar, and Monika Sachdeva.
\newblock The use of artificial intelligence based techniques for intrusion
  detection: a review.
\newblock {\em Artificial Intelligence Review}, 34(4):369--387, 2010.

\end{thebibliography}

\end{document}